\documentclass[10pt,a4paper,twoside,dvips]{article}

\usepackage{vmargin,fancyheadings,multicol,ifthen,cite,
        epsfig,wrapfig,calc,dcolumn,doublespace}  
\usepackage{wgn2}

\begin{document}
\begin{WGNpaper}{twocol}{
\sectiontitle{Meteor science}
\title{Analysis of the SonotaCo video meteor orbits.}
\author{Peter Vere\v{s} and Juraj T\'{o}th\\{Faculty of Mathematics, Physics and Informatics, Comenius University, Mlynska Dolina, 84248 Bratislava, Slovakia.\\
Email: {\tt toth@fmph.uniba.sk}} } \abstract{Since 2007 the
Japanese video network provided significant amount of meteor data
observed by multi station video meteor network located in Japan.
The network detects meteors mostly up to +2 magnitude and is
probably the most accurate and largest freely accessible video
meteor database up-to-date. In this paper we present our analysis
on the qualitative aspects of the meteor orbits derived from the
multi station video observation and the separation of the shower
meteors from the sporadic background.}

}%
%
\section{Introduction}
The SonotaCo database of the meteor orbits consists of 38710
entries. Of those, 37\% were identified as shower meteors. Data
were taken by 35 video meteor stations during years 2007 and 2008
in Japan \cite{1}. The survey goal was to cover the entire year.
Each database entry is equivalent to the heliocentric orbit
derived from the multi station video observation. In addition to
he heliocentric orbit, the meteor is identified as shower of
sporadic meteor, characterized by the apparent position on the sky
plane, angular velocity, magnitude and derived physical
parameters, such as geocentric velocity, relative height of the
meteor trail above the surface, duration of the visible trail,
etc. All parameters were derived by the UFOAnalyzer software and
orbits derived by the UFOOrbit software, both made by SonotaCo.
The notable advantage of the database is the very similar camera
setup of the network stations (e.g. lenses and CCD TV camers) and
unique tool for astrometric and velocity reduction, which almost
lacks individual observer influences. This makes the database very
homogenous.

\section{Database reduction}
In order to separate hight quality orbits, we set multiple
constraints on the database. The constrained parameters are
presented in the parentheses. Usually we adopted quality
determination according to Q3 condition for the high precision
computation (internal set of parameters for UFOOrbit). The most
important, the entire meteor trail had to be inside the field of
view of at least two video meteor stations (inout=3). Astrometric
accuracy and velocity determination drop with the observed trail
length, therefor the meteor trail had to be longer than 1 degree
($Q_{o}>1$) and the duration of the trail was over 0.3 seconds
($dur>0.3$). There parameters were set with respect to the network
camera setup. This provides at least 10 positions and velocity
measurements per meteor trail. Also the parameter $Q_{c}$ (cross
angle of two observed plains) had to be larger than 20 degrees.
The apparent velocity and derived velocities from two stations may
differ, our constrain allows the difference less than 10\%
($dv12\%<10$). One trail observed from two stations must be
detected to reach at least 50\% overlap (Gm\%) and ground
projection of the same meteor observed and derived for two
different stations must not have higher deviation than 0.1
degree(dGP). Finally, the total quality assessment parameter
larger than 0.7 (QA).

Number of meteor orbits that fulfill quality constraints is 8890.
47\% are meteors identified as shower meteors (IAU established
meteor showers and showers from the IAU working list).  292
meteors are on hyperbolic orbits ($a<0$ and $e>1$), of those 144
are sporadic and 148 were assigned to a meteor shower (mostly
Perseids, Orionids, Leonids, Dec. Monocerotids, sigma Hydrids).

Three-step algorithm of meteor shower identification by SonotaCo
is following. Particular meteor must be observer during the known
meteor shower activity (J6 catalog defined, \cite{1}) plus 10 days
variation. The backtraced meteor trail must lie within 100\% of
known meteor radiant. The geocentric velocity must be within 10\%
of the known mean shower geocentric velocity.

\section{Shower meteors identification}
Asignment of a meteor to a meteor shower is not a trivial task. In
our analysis, we employed orbit similarity criteria to distinguish
shower meteors from the non-shower component of the SonotaCo video
meteor database. Particularly, Southworth-Hawkins D-criterion
($D_{SH}$) was used for selected meteor showers \cite{2}.
Considering individual behavior of meteor stream orbits in
comparison to the mean orbit, we calculated the distribution of
D-criterion for Perseids (reference mean orbit by \cite{3}),
Orionids \cite{3}, Geminids \cite{4}, Leonids \cite{3}, sigma
Hydrids \cite{5} and Southern delta Aquarids \cite{6}. Histogram
of D-criterion of mentioned meteor showers separated from all
meteors (independently from the UFOOrbit identification of meteor
showers) are presented on Figure 1. The boundary D-criterion for
particular shower was derived from the point where the
distribution of D-criterion became eventually dispersed in the
sporadic background (dashed line in the plot of Figure 1 and 2).
If the meteor has a lower value of the specific D-criterion we
consider it as a shower meteor. Finally, we compared how many
particular shower meteors belong to the 8890 sample according to
the method by UFOOrbit and D-criterion. According to D-criterion,
some of shower meteors (by UFOOrbit classification) do not belong
to the meteor shower and on the contrary, some sporadic meteors
(by UFOOrbit) belong to the meteor shower but only in few cases.
Results are presented in Table 1.

\begin{table}[t]
\small
\begin{center}
\caption{Meteor showers identification according to UFOOrbit
algorighm and Southworth-Hawkins D-criterion. Shower-name,
$D_{SH}$-found limit for certain meteor shower identification,
$All<D_{SH}$-shower meteors derived according to D-criterion from
the entire subset of data (shower and non-shower), \%-percentage
of shower meteors in the shower component according to UFOOrbit
that did not fulfill D-criterion, Data-number of shower meteors
identified by UFOOrbit, Non-sporadic meteors according to UFOOrbit
belonging to the shower according to D-criterion.}
\begin{tabular}{l|ll|lll}
\hline
&Our data& & &SonotaCo&  \\
\hline
Shower & $D_{SH}$ & All $< D_{SH}$ &  $\%$ & Data  &  Non \\
\hline
PER                & 0.30 & 907 & 3.5   & 931 & 9\\
ORI                & 0.20 & 408 & 8.8   & 416 & 29\\
GEM                & 0.20 & 881 & 3.9   & 916 & 1\\
LEO                 & 0.20 & 90 & 15.2 & 105  & 1\\
HYD                 & 0.30 & 200 & 11.2  & 215 & 9\\
SDA                 & 0.15 & 103 &  2.0 &  104 & 1\\
\hline
\end{tabular}
\end{center}
\end{table}

\begin {figure}
\centerline{\includegraphics[width=6.0cm]{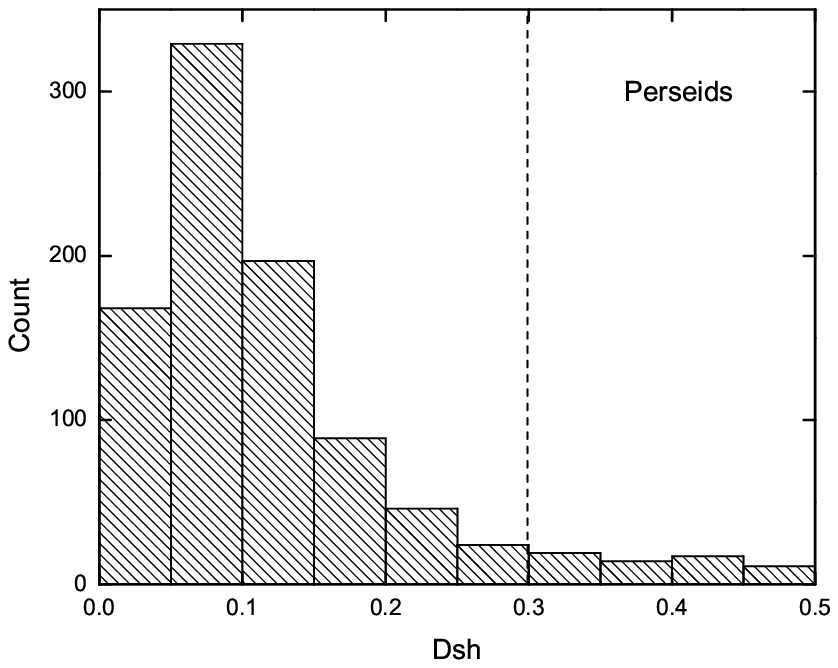}}
  \centerline{\includegraphics[width=6.0cm]{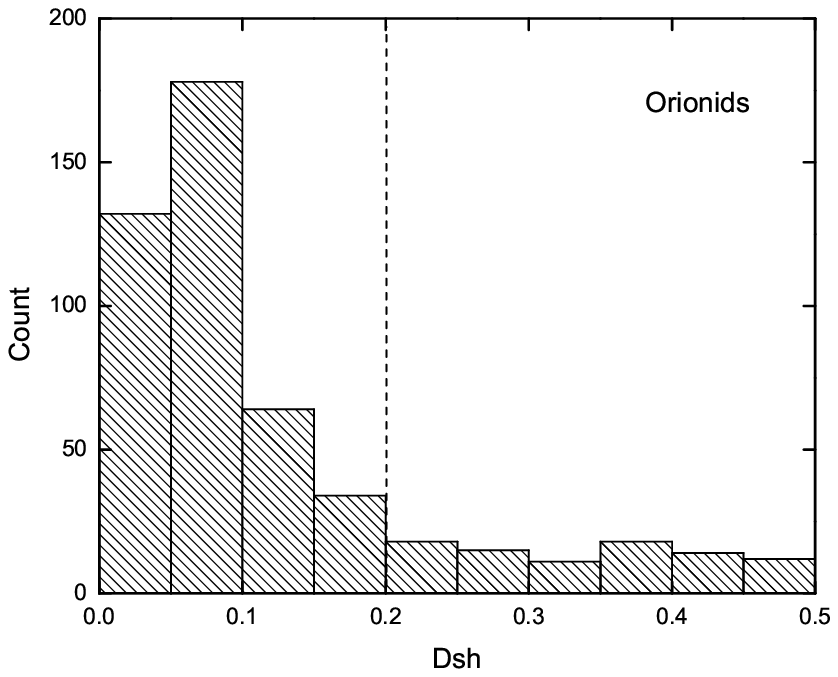}}
 \centerline{\includegraphics[width=6.0cm]{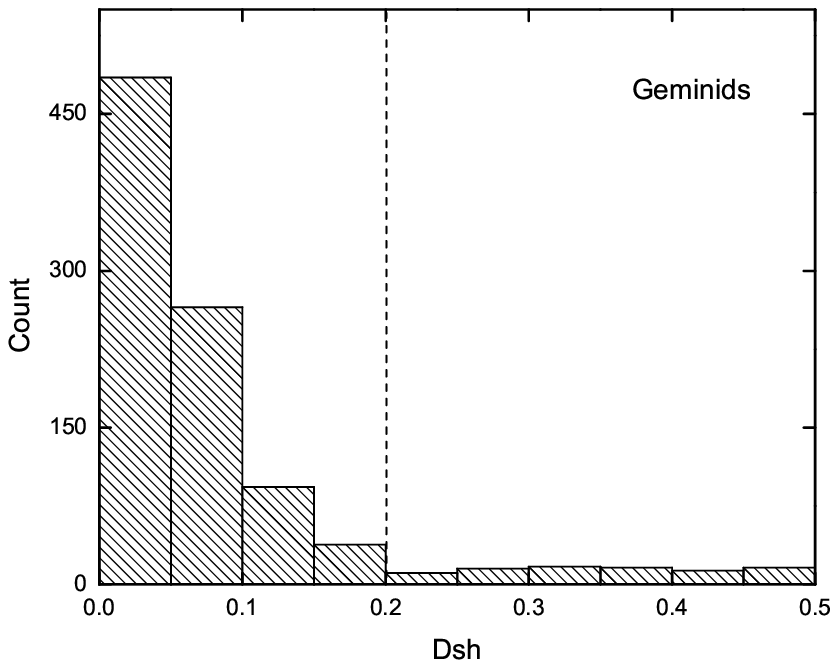}}

 \centerline{\includegraphics[width=6.0cm]{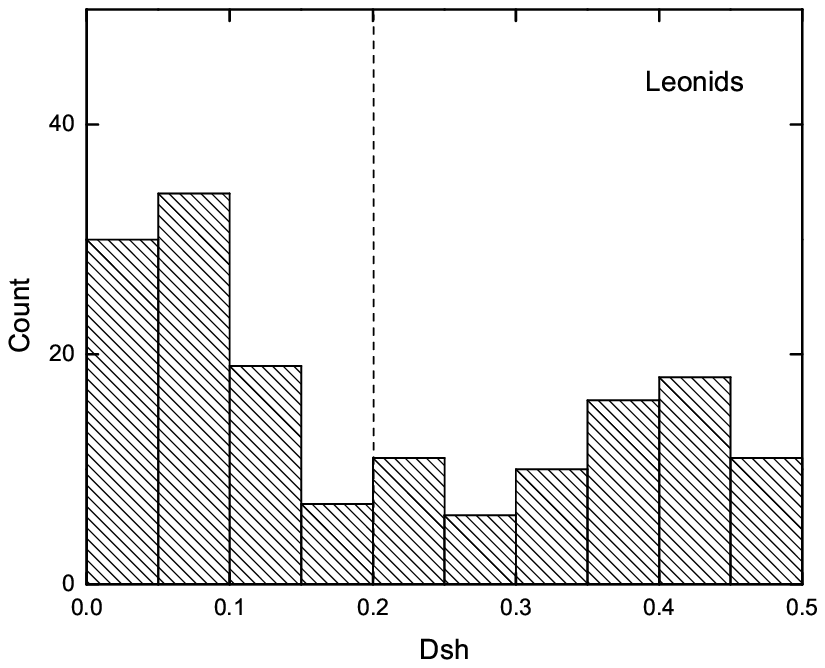}}
 \centerline{\includegraphics[width=6.0cm]{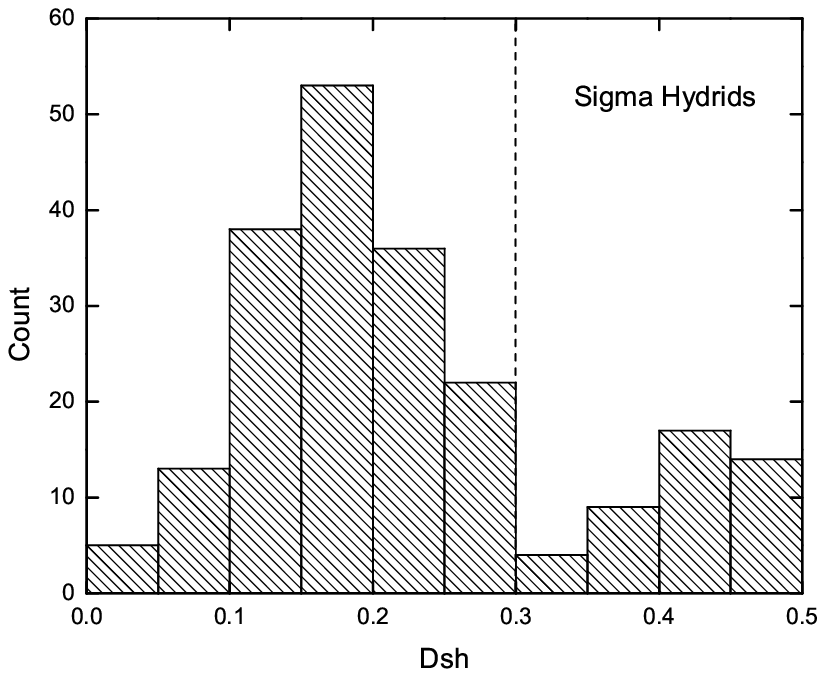}}

\caption{Southworth-Hawkins D-criteria for shower meteors from the
reduced database. Dashed line represents the limit that we adopted
to distinguish shower meteor from the sporadic background.}
\label{Figure 1}
\end{figure}
\begin {figure}

  \centerline{\includegraphics[width=6.0cm]{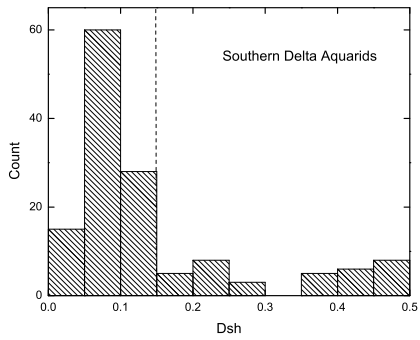}}
\caption{Southworth-Hawkins D-criteria for SDA. Dashed line represents the limit that we adopted
to distinguish shower meteor from the sporadic background.}
\label{Figure 2}
\end{figure}

\begin {figure}
\centerline{\includegraphics[width=8.0cm]{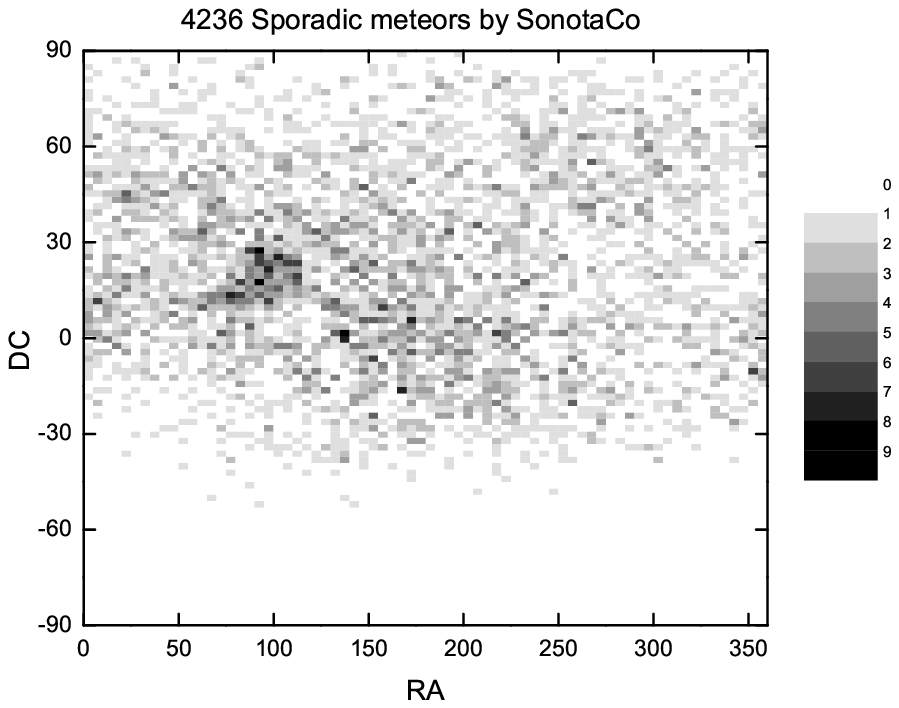}}

 \centerline{\includegraphics[width=8.0cm]{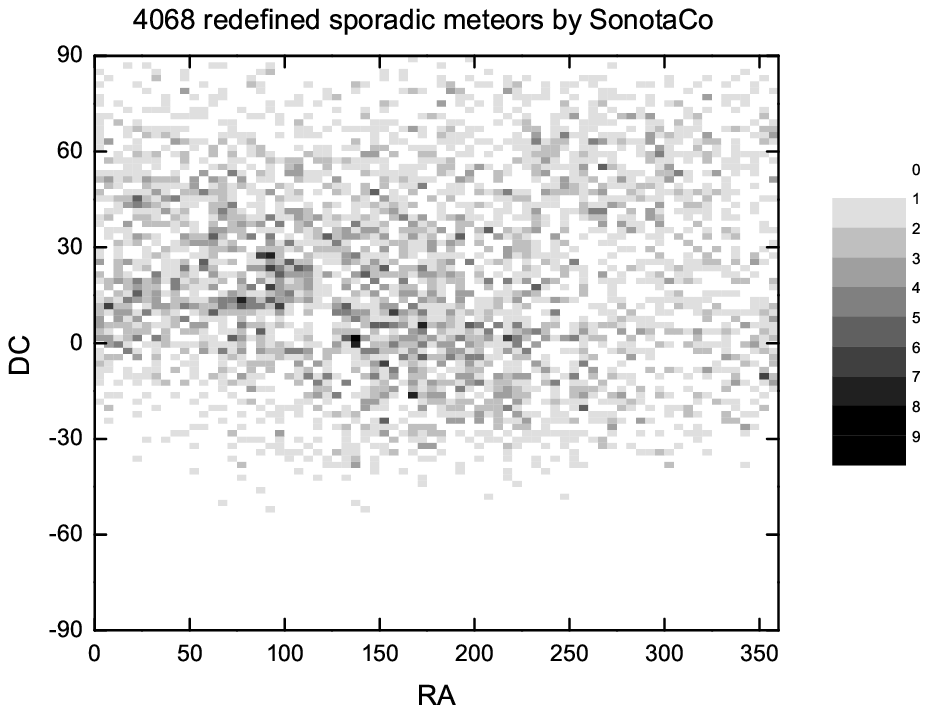}}

\caption{Density plots of sporadic population radiants from the
reduced UFOOrbit orbit database (left) and corrected sporadic
population - strong meteor shower members were separated using
D-criteria.} \label{Figure 3}
\end{figure}

\begin {figure}
\centerline{\includegraphics[width=8.0cm]{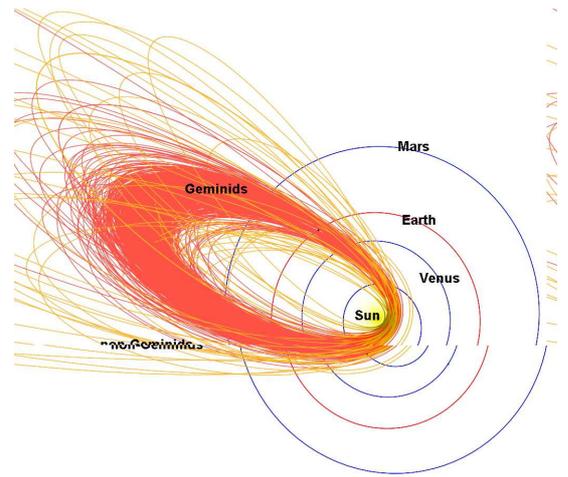}}
\caption{Orbits of Geminids meteors derived by the UFOOrbit
algorithm. Non-Geminids were identified as Geminids by UFOOrbit
but did not fulfill D-criterion for orbital similarity and are
apparently displaced from the standard meteor stream.}
\label{Figure 4}
\end{figure}

\begin {figure}
\centerline{\includegraphics[width=7.0cm]{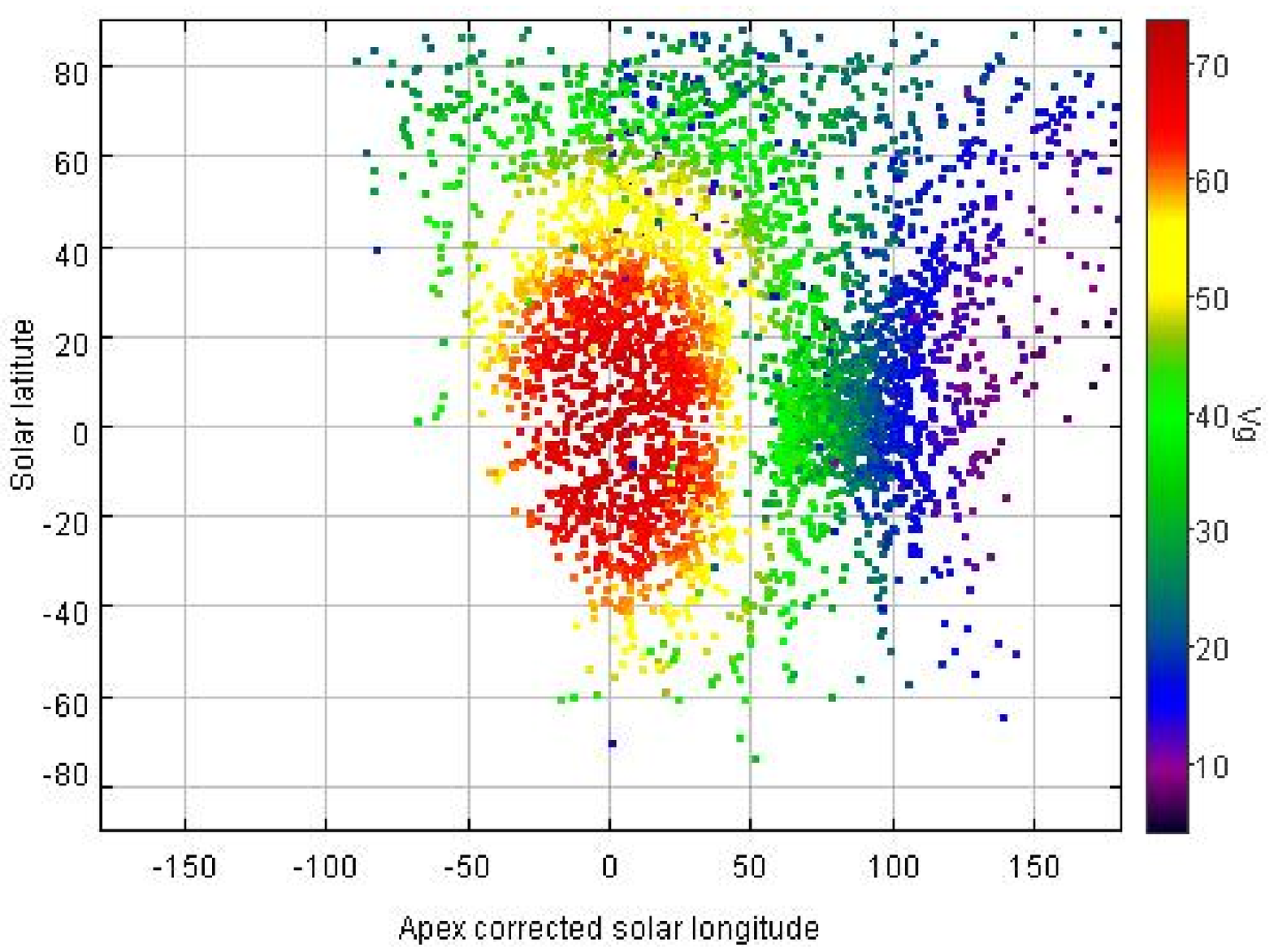}}
\centerline{\includegraphics[width=7.0cm]{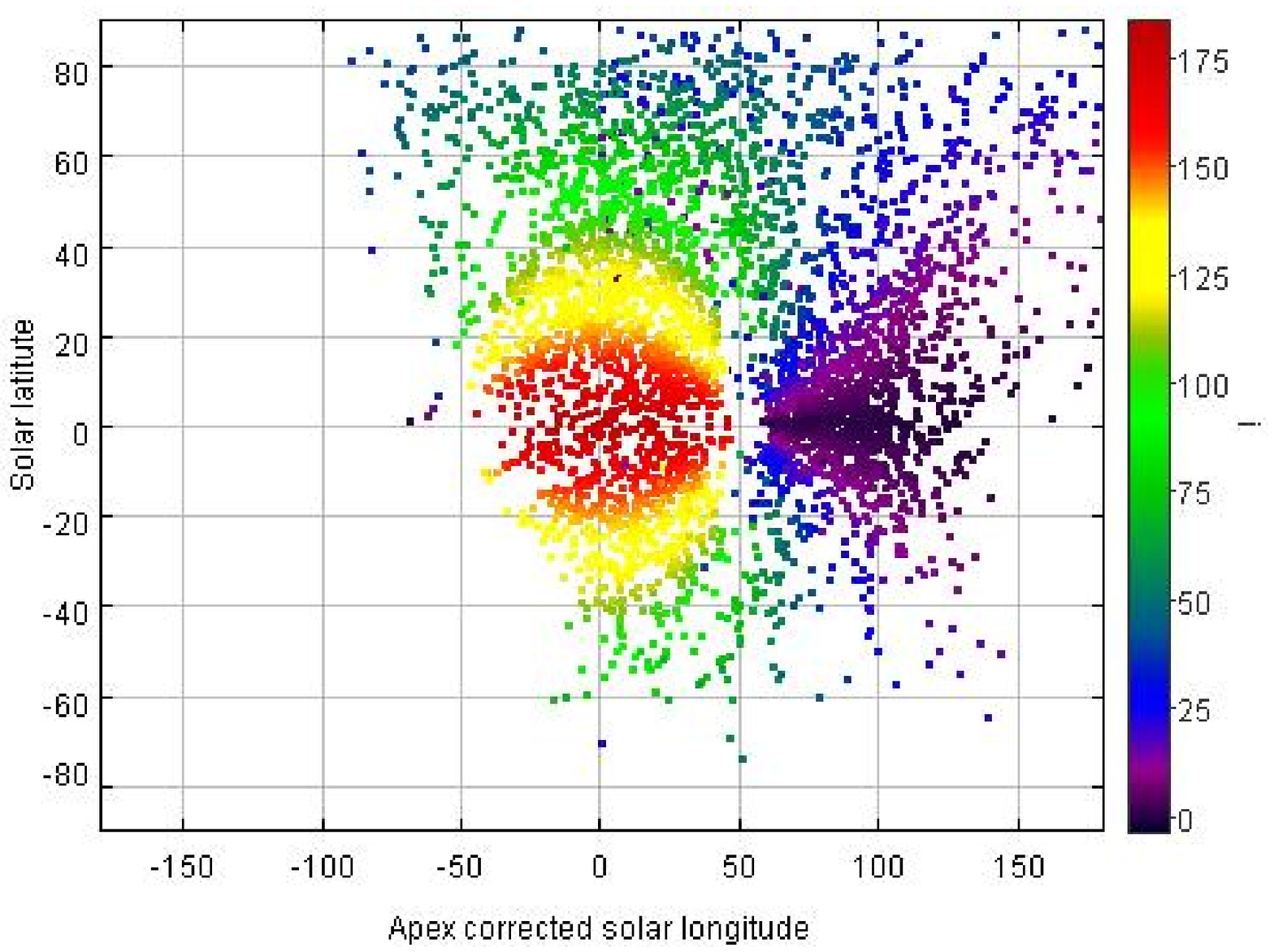}}
\caption{Earth apex corrected ecliptical coordinates of sporadic
meteor radiants. Color palette scale represents geocentric
velocity distribution, orbit inclination respectively.}
\label{Figure 5}
\end{figure}

Although 47\% of 8890 meteors are sporadic meteors according to
UFOOrbit classification, our investigation on 6 meteor showers
implies that the sporadic population in the database is
contaminated by shower meteors in a very small number (see Table
1, column Non). To obtain a rough estimate of the sporadic meteor
population, we applied Southworth-Hawkins D-criterion equal to
0.25 for 16 major showers that may have the most significant
contribution to the sporadic background of the SonotaCo database.
We used reference mean orbits of these meteor showers:
Quadrantids, Lyrids, pi Puppids, eta Aquarids, Arietids, sigma
Hydrids, June Bootids, Southern delta Aquarids, Perseids,
Draconids, Orionids, Southern Taurids, Northern Taurids, Leonids,
Geminids and Ursids (mean orbits from the photographic data,
\cite{5}). Radiant positions after the first separation procedure
are plotted in the density graph on Figure 3. We examined the
higher density of radiants on solar longitudes 265$^\circ$ $\pm$
30$^\circ$ ($\alpha=75^\circ - 115^\circ$, $\delta=10^\circ -
28^\circ$) and considered it as a contamination from the Taurid
complex (the position of the clump was similar as if Taurids were
active longer, meteors have similar geocentric velocities and
orbits). To separate assumed Taurid complex contamination, we used
Steel D-criterion equal to 0.2 for the mean orbit of the Southern
and Northern Taurids \cite{6}. This criterion is not sensitive to
the argument of the perihelion and the ascending node, therefor it
removes similar orbits even when the meteor was observed beyond
the established activity period. Finally, the sporadic meteor
count was derived to 4068. All year long activity is plotted on
Figure 3. There are two visible sources of sporadic meteors on the
apex corrected radiant distribution in the ecliptical coordinates
(Figure 5). The apex source contains meteors with high geocentric
velocities, orbits with high inclination and eccentricity. On the
contrary, antihelion source contains slow meteors with moderate
eccentricities and low inclinations. We may assume that meteors
from apex and toroidal sources have cometary origin and meteor
from antihelion source come from Near Earth Asteroid source.

\section{Conclusion}
The database of video observed meteors by SonotaCo contains
meteors that are relatively well distinguished as shower or
sporadic meteors among the high quality subset of data. For
further analysis of the meteor membership to the particular shower
we recommend to use additional tools for shower identification
such as orbit similarity D-criteria and orbital evolution with
respect to the mean reference orbit of the shower and the assumed
parent body. Meteors that were misidentified as shower meteors for
several examined meteor showers represent only small numbers of
the shower group identified by UFOOrbit. Separated sporadic
meteors demonstrated expected sky plane distribution in respect to
the Earth apex with the exceptional denser region which might be a
part of the wide Taurid complex. After all, the subset of video
meteor orbits we selected provides reliable data for both shower
and sporadic meteors.

\paragraph{Acknowledgment}
This work was supported by VEGA grant No. 1/0626/09 and Comenius University Grant No. UK/366/2009.
We are thankful to SonotaCo for his first review of the article and valuable corrections.

{}

\end{WGNpaper}
\end{document}